\documentclass[prl,floatfix,twocolumn,showpacs]{revtex4}
\usepackage{amsmath}
\usepackage{amsfonts}
\usepackage{mathrsfs}
\usepackage{epsfig}
\usepackage{graphicx}
\usepackage{dcolumn}
\usepackage{amsmath}

\def\be{\begin{equation}}
\def\ee{\end{equation}}

\begin{document}

\title{Emergence of universal scaling in financial markets from
mean-field dynamics}
\author{S.~V. Vikram$^1$ and Sitabhra Sinha$^2$}
{\affiliation{%
$^1$ Department of Physics, Indian Institute of Technology Madras,
Chennai - 600036, India.\\
$^2$ The Institute of Mathematical Sciences, C. I. T. Campus,
Taramani, Chennai - 600 113, India.}
\date{\today}
\begin{abstract}
Collective phenomena with universal properties have been observed in
many complex systems with a large number of components.
Here we present a microscopic model of the emergence of
scaling behavior in such systems, where the interaction dynamics
between individual components
is mediated by a global variable making the mean-field description
exact. Using the example of financial markets, we show that
asset price can be such a global variable with the critical role
of coordinating the actions of agents who are otherwise independent.
The resulting 
model accurately reproduces empirical properties such
as the universal scaling of the price fluctuation and volume
distributions, long-range correlations in volatility and multiscaling.
\end{abstract}

\maketitle

\newpage

Universal scaling behaviour is an emergent property of 
many complex  systems~\cite{yakovenko09}. In such systems,
the interactions between a large number
of individual components  yields
macro-scale collective behavior with features
that are almost invariant across different spatial and temporal
scales~\cite{farmer05}. A financial market provides a general and  useful paradigm 
of such a system, since it involves a large number of agents
whose actions are subject to internal and external influences, such as 
information about the state of the market as provided by
market indices~\cite{arthur99}. Despite this complexity, the availability
of a large volume of high-quality data for analysis has enabled 
the identification of  well characterized
statistical properties~\cite{kiyono06,bouchaud05}. 
These properties, including the fat-tailed
distribution of relative price
changes~\cite{mantegna95,gopikrishnan98} and intermittent
bursts of large fluctuations that characterize
volatility clustering~\cite{mandelbrot63}, appear to be universal: they are 
invariant across different markets, types of
assets traded and periods of observation~\cite{gabaix03}. More generally,
the question of how universal features emerge
from collective behavior in systems with many components
is not restricted to the purely economic domain. Thus, 
new approaches to understanding the behavior of financial markets
may contribute to the understanding of  the physics of non-equilibrium 
steady states in general.

Mainstream economic theories for price fluctuations
of financial assets typically assume the {\em efficient market
hypothesis}~\cite{fama70}.
According to this, price variations reflect changes in
the fundamental (or ``true") value of the assets. However, detailed analysis of
data from actual markets show that much of the observed price
variation cannot be explained solely in terms of changes in economic
fundamentals~\cite{shiller87}. The absence of a strong correlation
between large market fluctuations and purely economic factors 
leaves unresolved the question of why markets are so volatile. As the
dynamics of markets are a result of the collective behavior of many
interacting constituents, models based on statistical physics have
been proposed to explain
the observed universal behavior~\cite{lux99,eguiluz00,krawiecki02,daniels03}.
Most such models consider explicit interactions between agents to
reproduce a very limited set of the universal empirical features.
However, it is possible that the observed complex
behavior is a result of a mean-field-like global variable mediating
the dynamics of components, which are therefore coupled only
indirectly.
Such a potential simplification in analyzing the non-equilibrium
steady state for markets holds promise as a general descriptive
framework for the dynamics of many complex systems.

In this paper, we present a model for market dynamics
where the action of each agent is governed solely by 
global information about the system, viz., the price $p_t$ of the
single asset being traded. At each time-step, every agent goes through
a two-step stochastic process, analogous to decision
making in an uncertain environment. 
Based on the deviation of the
instantaneous price from its long-term average (representing the
notional fundamental value of the asset) and the direction of price
movement, each agent decides
(a) whether to trade, and (b) if yes, whether to buy or sell.
The price in turn evolves as a function of the net demand,
measured as the difference between the number of buyers and sellers.
Thus, our approach falls broadly within the theoretical framework that
treats markets as a system of spins, but it differs from earlier models 
in not having direct Ising-like interactions between the
agents~\cite{krawiecki02,chowdhury99}. Further, the fluctuations in
the model variables are endogenous to the system and are not
responses to
external noise simulating the arrival of news or
information~\cite{lux99}.
Despite its simplicity the model
reproduces the observed universal
properties of markets.
These include the scaling behavior of the distribution of price
fluctuation measured by the relative logarithmic change, viz., 
the return, $R_{t,\Delta t} = \ln (p_{t+\Delta t}/p_t)$ defined over a
time-interval $\Delta t$. The cumulative distribution of $R_{t,\Delta
t}$ shows a power-law tail
with a characteristic exponent $\alpha \sim 3$ for many different
markets -- a robust property referred to as the {\em inverse cubic
law}~\cite{gopikrishna98,lux96}. Our model, which displays power-law
scaling in the return distribution over a large region of the parameter
space, can quantitatively reproduce the inverse cubic law on
introducing heterogeneity among the agents.
For the same parameters,
the scaling exponent $\zeta_V$ for the distribution of trading
volume in a given interval
of time, $V_t$, agrees with the empirical values reported in
Ref.~\cite{gopikrishna00}.
Moreover, the time-series generated by the model exhibits multifractal
statistics~\cite{ghashghaie96} and the auto-correlation of absolute
returns decay slowly, a
signature of volatility clustering 
seen in actual markets~\cite{muzy00}.
We also give an analytical derivation of the relation between the
scaling exponents for return and volume distributions generated by the model, 
which is in good agreement with the empirical literature.

We consider the market to comprise $N$ agents, each of whom are in one
of three possible states at time $t$, viz., $S_i (t) = 0$ (not trading), $+1$
(buying) and $-1$ (selling) ($i = 1,\ldots,N$).
For simplicity, we assume that an agent can trade a unit quantity of
asset at a given instant. The change in the price of the asset is
driven by the net demand, as measured by the global order parameter $ M_t
= \sum_{i} S_i(t)/N$. Thus, after time instant $t$, the asset price
changes to $p_{t+1} = [(1 + M_t)/(1 - M_t)] p_t$, with $p_0 > 0$,
which ensures that the price is always positive and
rises (falls) when relatively more agents buy
(sell) it, with the inactive agents, i.e., $S_i(t)=0$, not affecting the process. 
A price equilibrium ($p_{t+1} = p_t$) is achieved when supply equals
demand ($M_t=0$), while in
extreme cases, when all $N$ agents buy (sell), the
price diverges (crashes to 0).
We have verified that the exact form of the price function is not
critical to obtain the results described here.

In our model, the net demand $M_t$ is driven by the 
collective behavior of agents, with each individual's state $S_i$
in turn evolving as a result of fluctuations in the instantaneous price
$p_t$ around the asset's fundamental value $p^*_t$ as perceived by an
agent. As the ``true" value of $p^*_t$ is privileged information and
therefore inaccessible to an agent, it is estimated based on the
observed price time-series
as $p^*_t \simeq \langle p_t \rangle_\tau$, the long-time moving
average measured over a window of duration $\tau$ ($=10^4$ time units
for the results shown in the paper). Note that
previous studies have shown that several features of empirical market
dynamics are determined by an effective potential
defined in terms of the long-term moving average of
price~\cite{alfi06}. Given the price information, an agent $i$ decides to
trade at time $t$ according to the probability
\begin{equation}
P[|S_i(t)|=1] = 1-P[S_i(t)=0] = \exp\left({- \mu \left|\log{\frac{p_t}{\langle 
p_t \rangle_{\tau}}}\right|}\right),
\label{trade_or_not}
\end{equation}
where the parameter $\mu$ is a measure of the sensitivity of an agent
to the magnitude of deviation of the price from its perceived
fundamental value.
For $\mu = 0$, the system reduces to a 2-state model where every agent 
trades at all time instants.

Once an agent has decided to trade at time $t$, it still has to choose
whether to buy [$S_i(t)=+1$] or sell [$S_i(t)=-1$]. Using the simple assumption that this is a
random process, we allow each trader to either buy or sell with equal
probability, independent of the price movement. We have verified that 
introducing more complicated rules 
based on consideration of supply and demand, where the decision to
buy or sell depends on the 
instantaneous price
fluctuations (e.g., as measured by the return), do not 
qualitatively change the results reported here.

\begin{figure}
\centering
\includegraphics[width=0.99\linewidth,clip]{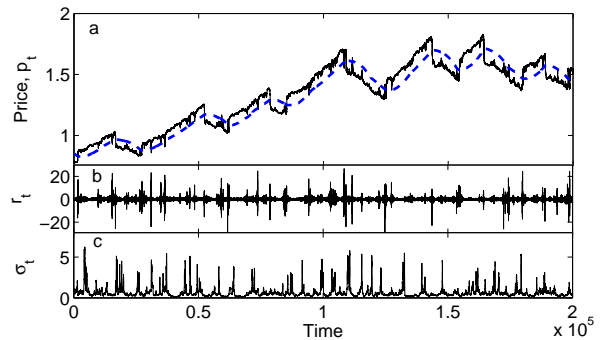}
\caption{{\bf Time evolution.}
  (a) The time-series of price $p_t$ (solid line) and 
  its moving average $\langle p_t \rangle_\tau$
  (broken line). The corresponding
  (b)~normalized returns, $r_t$,
  and (c)~volatility,
  $\sigma_t$, calculated as the standard deviation of returns
  in a moving window of
  interval $\delta t = 100$, show intermittent bursts of large price
  fluctuations. This indicates the presence of volatility clustering.
  Varying $\delta t$ does not change the result qualitatively.
  The model parameters are $N =2 \times 10^4$ and $\mu = 100$.}
\label{timeseries}       
\vspace{-0.15cm}
\end{figure}

Time-evolution of the asset price, $p_t$, shown in
Fig.~\ref{timeseries}~(a), is qualitatively similar to the
time-series of stock prices or indices observed in real markets.
The moving average of $p_t$ (broken line) which is the agents'
perceived fundamental value of the asset, tracks a smoothed pattern
of price fluctuations coarse-grained over a time-scale $\tau$ corresponding
to the size of the averaging window. The
normalized returns $r_t$ for $\Delta t = 1$ time unit, obtained
from $R_t$ by subtracting the mean and dividing by the standard deviation of the
entire return time series,
exhibits significantly large deviations relative to that expected from
a Gaussian distribution [Fig.~\ref{timeseries}~(b)]. These
intermittent bursts of large fluctuations have a tendency to aggregate
together. 
This is seen more clearly from the volatility $\sigma_t$, 
which is a measure of risk (the
unpredictable change in the value of an asset) and may be calculated as
the standard deviation of $r_t$ over a moving window. The clustering
of volatility seen in 
Fig.~\ref{timeseries}~(c) is 
a universal feature of financial markets.

The nature of price fluctuations can be examined in more detail
by focusing on the cumulative distribution $P_c$($r_t > x$). When the
agents are homogeneous (i.e., having the same sensitivity $\mu$), 
this distribution shows power-law tails having exponent $\alpha \simeq
2$ for a large range of values of $\mu$ (viz. $\mu > 50$).
For lower values of $\mu$ ($< 10$) the distribution is exponential.
\begin{figure}[agent_dist]
\centering
\includegraphics[width=0.99\linewidth,clip]{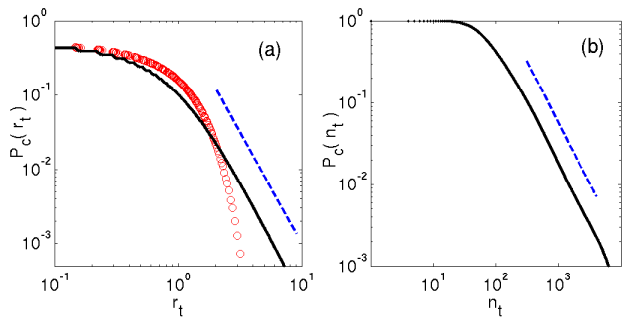}
\caption{{\bf Distributions.} (a) Cumulative distribution of normalized returns for
heterogeneous agents with $\mu$ distributed uniformly over $[10,200]$.
The broken line indicates a power-law exponent of $-3$ and the circles
represent the standard normal distribution.
(b) The corresponding cumulative distribution of the number of agents 
trading at a given time instant $t$, $n_t$, with the broken line 
indicating a power-law exponent of $-1.5$. The results are obtained by
averaging over multiple realizations of the model with
$N = 10^4$ agents simulated over $2 \times 10^5$ time units.}
\label{agents_fig}       
\vspace{-0.15cm}
\end{figure}
In reality, agents will differ in their responses to the same 
stimulus. This heterogeneity in agent behavior is modeled by 
a distribution of the sensitivity parameter $\mu$ that
measures the degree of risk-aversion in an individual.
Fig.~\ref{agents_fig}~(a) shows that the cumulative distribution for $r_t$
quantitatively reproduces the inverse cubic law ($\alpha \simeq 3$) 
when $\mu$ for each agent is randomly selected from an interval. 
To accurately determine the numerical value of the return exponent
$\alpha$, we use the Hill estimator, $\gamma_{k,n}$, for a time-series
of length $n$, whose inverse approaches the true
value of $\alpha$ as the order statistic $k \to \infty$ with
$\frac{k}{n} \to 0$~\cite{hill75}. To avoid the bias arising from
finite length of the time-series, we have used a subsample bootstrap
method to estimate the optimal $k$~\cite{pictet98}. 
Using this method, the estimated value of the exponent $\alpha$ is
3.11 for the positive tail of the return distribution [shown in
Fig.~\ref{agents_fig}~(a)] and 3.12 for the negative tail. 
We have verified that this long-tailed behavior of returns is robust 
with respect to variations in the interval and the nature of the
distribution for $\mu$.

As the model assumes that each trading agent buys or sells a unit 
quantity of the asset, the total number of traders at any instant $t$,
viz., $n_t = \sum_i |S_i (t)|$, is equivalent to the trading 
volume $V_t$. The distribution of this variable also exhibits a
power-law scaling, with the exponent $\zeta_V \simeq 1$ when the agents
are homogeneous. The heavy-tailed nature of $n_t$ distribution is even
more robust than that of $r_t$, as we observe a power-law tail also for
lower values of $\mu$ 
(where the return distribution is exponential).
On introducing heterogeneity among agents as explained before, the
cumulative distribution of $n_t$ is seen to be a power-law, whose
exponent is evaluated by the Hill estimator to be
$\zeta_V \simeq 1.63$ (using the same parameters for which $\alpha
\simeq 3$) [Fig.~\ref{agents_fig}~(b)]. This is almost identical to
the trading volume exponents reported for different
markets~\cite{plerou07}. In order to check the sensitivity of our
results on the 
assumption that an agent can trade only a unit quantity, 
we have verified that a Poisson distribution of the number of units
traded by an agent does not change the results qualitatively.
Thus, our model suggests that heterogeneity in agent behavior is a
key factor for explaining the quantitative properties of 
the observed distributions.
It implies that when the behavior of agents become more homogeneous,
e.g., during a market crash, the return exponent $\alpha$ will
tend to decrease. This is intriguing in light of earlier 
work~\cite{kaizoji06} showing that
the power-law exponent for
the distribution of relative prices during a crash has a significantly
different value from that seen at other times.

\begin{figure}[volatility]
\centering
\includegraphics[width=0.99\linewidth,clip]{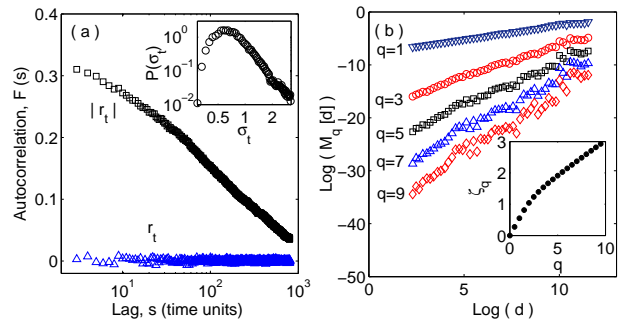}
\caption{{\bf Correlations.} 
  (a) The auto-correlation of $r_t$ (triangle) rapidly falls
  to noise level but its absolute value (squares) shows long-term
  memory. (Inset) Probability distribution of the volatility, $\sigma_t$. 
  (b) The $q$-th moments of absolute value of the fluctuations in
  return have a power-law scaling relation with respect to the
  time-scale $d$. The inset shows the
  non-linear variation of the corresponding power-law exponent, 
  $\zeta_q$, indicating multifractality. 
  The model parameters are $N = 2 \times 10^4$ and $\mu = 100$.}
\label{volatility_fig}
\vspace{-0.15cm}
\end{figure}
Turning now to the correlation properties of the return
time-series, we see that $r_t$ is uncorrelated, as expected from the
efficient market hypothesis~\cite{fama70}. However, the absolute
values, which are a measure of the volatility, show a slow
logarithmic decay in their auto-correlation
(Fig.~\ref{volatility_fig}~(a)), which is a signature of
long-memory effects operating in actual markets~\cite{muzy00}.
Fig.~\ref{volatility_fig}~(a) (inset) shows the bulk of the volatility
distribution which has a log-normal form as found
empirically~\cite{cizeau97}.
To understand better the temporal organization of price fluctuations
than is possible with the 2-point correlations considered above, we
consider the $n$-point correlations as reflected in the multifractal 
spectrum~\cite{matia03,baldovin07}. 
Fig.~\ref{volatility_fig}~(b) shows the
power-law scaling of the $q$-th moment  $M_q (d)$ of the absolute value of
fluctuations as a function of the time scale ($d$) being considered.
The resulting power-law exponents $\zeta_q$ do not have a simple
linear relation to $q$ (inset), indicating that the process is
multifractal. Thus, our model also reproduces the multifractal
nature of financial markets~\cite{ghashghaie96,muzy00}. 

The genesis of the power-law scaling relations in the model is
strongly connected to the dynamics by which agents decide to trade,
which results in a variable number of agents buying/selling at a given
instant. This is illustrated by the absence of power-law scaling when 
the number of trading agents do not change with time 
(viz., $\mu = 0$ for which $n=N$).
Further, it is the long-tailed nature of the distribution
of $n_t$ that is responsible for the heavy tails of the returns. For
instance, imposing a log-normal form on $n_t$ rather than generating
it by
using Eq.~(\ref{trade_or_not}), again results in fat tails for
$r_t$. This dependence of the long-tailed nature of the returns on
the distribution of number of trading agents can be
analytically derived as follows.
First, we note that, if 
the number of trading agents is a constant ($n_t = n$), the returns
follow a Gaussian distribution with mean 0 and variance,
$\sigma^2 \sim n$. 
Therefore, when the number of traders changes over time, 
with $n_t$ following a distribution $P (n)$,
the corresponding return distribution $P(r)$ can be expressed as a sum over 
many conditional distributions $P(r|n)$:
\begin{equation}
P(r)= \sum_{n=1}^N P(r|n) P(n) = \sum_{n=1}^{N} \frac{1}{\sqrt{2 \pi
n^2}}{\rm exp}(-r^2/2 n) P(n),
\label{eq1}
\end{equation}
where $N$ is the maximum number of agents who can trade.
If the cumulative distribution for $n_t$ follows a power law with
exponent $\zeta_V$ as obtained from our model, Eq.~(\ref{eq1}) can
be rewritten as
$ P(r) \sim [1/\sqrt{2 \pi}] \sum_{n=1}^{N} n^{-(\zeta_V + \frac{3}{2})} 
\rm{exp}(-r^2/2 n)$.
Replacing the sum by an integral and taking the upper limit $N \to
\infty$, we get a closed form solution
\begin{equation}
P(r) = C_{\zeta_v} ~K(0.5+\zeta_V,1.5+\zeta_V,-r^2/2),
\label{eq2}
\end{equation}
where $K(.)$ is the Kummer confluent hypergeometric function
and the normalization constant
$C_{\zeta_v} = \frac{\Gamma (0.5 + \zeta_V) \Gamma ( 1 +
\zeta_V)}{\sqrt{2 \pi} \Gamma (\zeta_V) \Gamma (1.5 + \zeta_V)}.$
Numerically evaluating $K(.)$ gives a power-law distribution for $r$.
For half-integral values of the exponent $\zeta_V$, Eq.~(\ref{eq2}) 
simplifies to a form where the power law
nature of the return distribution is evident. E.g., for $\zeta_V = 3/2$, 
as obtained in our model for heterogeneous distribution of agents,
$P(r) = C_{\zeta_V=3/2} (1/r^4) [4 - 2 \rm{e}^{-r^2} (2 + r^2)],$
with $C_{\zeta_V=3/2} \simeq 1.67$. For large $r$, $P(r) \sim r^{-4}$,
indicating that the cumulative distribution of returns will have a
power-law tail with exponent $\alpha = 3$ (i.e., the 
inverse cubic law).

In this paper we have presented a model for the dynamics of complex systems 
which
quantitatively reproduces the observed universal properties of markets without
considering explicit interactions among agents or prior assumptions about
individual trading strategies (e.g., chartists vs.
fundamentalists)~\cite{lux99}. Recent work on other aspects of
financial markets have shown that coherent collective behavior can
emerge in a system through components responding to the same global
signal~\cite{pan07}. We show that the price of an asset can play the role
of such a mediator that generates effective interactions between agents,
resulting in a non-equilibrium steady state
characterized by scaling distributions.
Heterogeneity of agent behavior is seen to be critical for obtaining
the inverse cubic law, suggesting that in normal circumstances agents
differ significantly in terms of their response to similar market
signals. On the other hand, when the agents are more homogeneous in
their behavior (as during a crash), the model exhibits even fatter
tails.
Possible extensions of our model include the introduction of a volume
dynamics that decides the quantity of assets traded by an agent at a
particular time instant, the inclusion of multiple assets and
considering the effect of
external news.
The framework presented here can be applied to many other complex
systems whose emergent phenomena can be explained in terms of indirect
interactions between components mediated by a mean-field-like variable.


We thank D.~Dhar, F.~Lillo, M.~Marsili, P.~Ray and P.~Shukla for 
helpful discussions. We are grateful to G. Menon and S. Sridhar for
numerous suggestions for improving the manuscript.
This work is supported in part by IMSc Complex Systems (XI Plan) Project.


\end{document}